\documentclass[%
reprint,
superscriptaddress,
nobalancelastpage,
 amsmath,amssymb,
 aps,
pra,
]{revtex4-2}
\usepackage{amsfonts}
\usepackage{graphicx}% Include figure files
\usepackage{dcolumn}% Align table columns on decimal point
\usepackage{bm}% bold math
\usepackage{lineno}
\usepackage{color}
\begin{document}

\preprint{APS/123-QED}

\title{Towards universal transformations of orbital angular momentum of a single photon}% Force line breaks with \\

%\affiliation{%

\author{Dong-Xu Chen}
 \email{chendx@sru.edu.cn}
 \affiliation{Quantum Information Research Center, Shangrao Normal University, Shangrao, Jiangxi 334001, China}
%% \altaffiliation[Also at ]{Physics Department, XYZ University.}%Lines break automatically or can be forced with \\
\author{Yunlong Wang}%
\affiliation{Shaanxi Key Laboratory of Quantum Information and Quantum Optoelectronic Devices, School of Physics of Xi'an Jiaotong University, Xi'an 710049, China}
\author{Feiran Wang}%
\affiliation{School of Science of Xi'an Polytechnic University, Xi'an 710048, China}
\author{Jun-Long Zhao}%
\affiliation{Quantum Information Research Center, Shangrao Normal University, Shangrao, Jiangxi 334001, China}
\author{Chui-Ping Yang}%
 \email{yangcp@hznu.edu.cn}
 \affiliation{Quantum Information Research Center, Shangrao Normal University, Shangrao, Jiangxi 334001, China}
 \affiliation{School of Physics, Hangzhou Normal University, Hangzhou, Zhejiang 311121, China}

%\begin{large}
\date{\today}% It is always \today, today,
\begin{abstract}
%\begin{large}
High-dimensional quantum systems offer many advantages over low-dimensional quantum systems. Meanwhile, unitary transformations on quantum states are important parts in various quantum information tasks, whereas they become technically infeasible as the dimensionality increases. The photonic orbital angular momentum (OAM), which is inherit in the transverse spatial mode of photons, offers a natural carrier to encode information in high-dimensional spaces. However, it's even more challenging to realize arbitrary unitary transformations on the photonic OAM states. Here, by combining the path and OAM degrees of freedom of a single photon, an efficient scheme to realize arbitrary unitary transformations on the path-OAM coupled quantum states is proposed. The proposal reduces the number of required interferometers by approximately one quarter compared with previous works, while maintaining the symmetric structure. It is shown that by using OAM-to-path interfaces, this scheme can be utilized to realize arbitrary unitary transformations on the OAM states of photons. This work facilitates the development of high-dimension quantum state transformations, and opens a new door to the manipulation of the photonic OAM states.
%\end{large}
\end{abstract}

\maketitle

%\tableofcontents
%\begin{large}
\section{Introduction}
High-dimensional (HD) quantum systems have attracted increasing attention in recent years \cite{bouchard2017high,PhysRevLett.123.070505,Daniele2019,erhard2020advances}. Compared with two-dimensional quantum systems, HD quantum systems are more advantageous. For instance, HD quantum systems allow high-capacity quantum communication protocols to be realizable, which benefits from their larger Hilbert spaces that can be used to encode much more information \cite{PhysRevLett.108.143603,Hu:20}. HD quantum systems can also enhance the security in quantum key distributions \cite{PhysRevLett.88.127902}, and show a larger violation of local realism \cite{PhysRevLett.104.060401,Weiss_2016}. Therefore, as a counterpart to two-dimensional systems, HD quantum systems have important potential applications in various fields, including HD quantum communication and HD quantum computation. 

Most research fields focus on two-dimensional systems, which use a qubit (i.e., a two-state quanton) as the basic information carrier. In contrast, the basic building block in HD systems is a qudit, which can be seen as a quanton that has more than two discrete states. As flying quantons, photons are considered as one of the most promising candidates for quantum communication due to their long coherence time. A photon has many degrees of freedom (DoFs), e.g., polarization, path \cite{PhysRevLett.125.230501,hu2020experimental}, time \cite{steinlechner2017distribution,PhysRevLett.118.110501,islam2017provably}, and orbital angular momentum (OAM) \cite{100100,10010,PhysRevLett.120.260502}. Each DoF can be used to encode information. The polarization DoF of photons is a binary with two distinct polarization modes; while in principle, the path, time and OAM DoFs of photons have infinite modes, which makes them natural carriers for encoding a qudit.

The OAM of a photon indwells in a twisted photon with a helical phase factor $e^{i\ell\varphi}$, where $\varphi$ is the azimuthal angle that is in the plane transverse to the propagation direction, and $\ell$ is an integer which, in principle, has infinite values. A photon with a phase factor $e^{i\ell\varphi}$ is said to carry an OAM $\ell\hbar$ \cite{PhysRevA.45.8185}. Compared with the path DoF, the photonic OAM has advantages in some aspects. For example, as an internal DoF, the OAM of a photon maintains a mutual stability when the photon travels in space. Thus, it is more suitable to use the photonic OAM in the air-to-air communication \cite{d2012complete,Willner:15,wang2019twisted}. The photonic OAM can be readily extended to an HD space without additional resources. To date, the generation and measurement of the photonic OAM have been well studied (see Refs. \cite{Yao:11,Padgett:17,erhard2018twisted} and references therein). However, only some optical elements have been designed to manipulate the OAM, e.g., the dove prisms \cite{PhysRevLett.88.257901}, q-plate \cite{PhysRevLett.96.163905} and spiral phase plate \cite{PhysRevA.103.052404}, etc. These elements can only realize specific operations on the OAM states; whereas as the dimensionality grows, it becomes challenging to implement some important transformations (e.g., X gate \cite{Schlederer_2016,PhysRevLett.116.090405,Chen_2017,isdrailua2019cyclic,PhysRevA.99.023825}, Fourier transformation \cite{PhysRevApplied.14.034036,PhysRevA.104.012413}) in HD quantum communication and HD quantum computation.

In various quantum information tasks, unitary transformations on quantum states are important parts, and they are generally realized by decomposing them into basic quantum gate operations which are easy to implement. For a qubit system, there exists complete sets of basic quantum gates, which can be used to perform a general unitary transformation on quantum states in a variety of platforms. For a qubit encoded via photonic polarizations, a combination of wave plates is sufficient to realize any operation on the qubit states \cite{PhysRevA.85.022323}. However, for a qudit, performing unitary transformations on the qudit states is more challenging as more resource is required, and it brings difficulty in the experimental implementation. 

The work on the universal implementation of unitary transformations on HD quantum states was pioneered by Reck \textit{et al}. \cite{PhysRevLett.73.58}. They decomposed arbitrary unitary HD transformations into the splitting operations between two adjacent modes. The architecture of the algorithm formed a triangular layout. Later, it was improved by Clements \textit{et al.} with a new symmetric rectangular design \cite{Clements:16} which is symmetric with respect to different paths, thus it is more robust to the noise introduced by the interferometers, and has a shorter optical path. Both approaches \cite{PhysRevLett.73.58,Clements:16} turn an HD unitary transformation matrix to a diagonal matrix by making the off-diagonal entries zero in sequence. Each step of elimination requires a splitting operation between two adjacent modes. Thus the total number of required splitting operations is $N(N-1)/2$, where $N$ is the dimensionality of the input quantum state. For a qudit encoded by photonic paths, the splitting operation can be realized by a Mach-Zehnder interferometer (MZI) with two tunable phase shifters. Therefore, the number of required MZIs scales identically as $N(N-1)/2$. On the other hand, the integrated photonic circuit is becoming a promising platform to realize linear optical experiments compared with the bulk optical elements, because of its reconfigurability, high stability, and versatility \cite{bogaerts2020programmable,9205209,chen2021,lu2021,al2022}. The elimination-based approaches are now the primary routes to implement HD transformations in integrated photonic circuits \cite{univer,PhysRevLett.123.250503,chi2022programmable,hoch2022reconfigurable}. 

In this work, by combining the path and OAM DoFs of a single photon, we propose a scheme that realizes arbitrary unitary transformations on the path-OAM coupled HD states. Compared with the proposals in \cite{PhysRevLett.73.58,Clements:16}, our scheme reduces the number of required MZIs by approximately one quarter while maintaining the symmetric structure, at the cost of additional dove prisms and a longer optical depth. Our scheme uses two OAM eigenstates in the OAM subspace, with the same OAM magnitude but opposite signs, to couple with the path DoF. The key element in our scheme is an OAM-dependent beam splitter which can realize different splitting operations on two different OAM states, thus incorporating two MZIs into a single MZI structure.

In principle, the decomposition of HD transformations into basic operations applies to any degree of freedom; while, for transformations on the OAM states, straightforwardly applying the elimination-based approaches is technically infeasible. In Ref.~\cite{PhysRevLett.119.180510}, it was demonstrated that HD X and Z gates, and integer powers of them, can form another complete set of HD quantum gates. By using the Heisenberg-Weyl operators, \cite{PhysRevLett.119.180510} decomposed an arbitrary HD transformation into sum of $N^2$ operations, each of which was comprised of HD X and Z gates, and integer powers of them. But the coefficients of the $N^2$ operations were not explicitly given in \cite{PhysRevLett.119.180510} and it is even more challenging to experimentally realize the $N^2$ operations. Here, we further show that by using OAM-to-path interfaces to redirect the OAM states onto different paths, our scheme can be utilized to implement arbitrary transformations on HD states enabled by the OAM DoF alone. Our approach provides an explicit and simpler implementation of unitary transformations on OAM states compared with \cite{PhysRevLett.119.180510}.

\begin{figure}[tbp!]
\centering\includegraphics[width=0.4\textwidth]{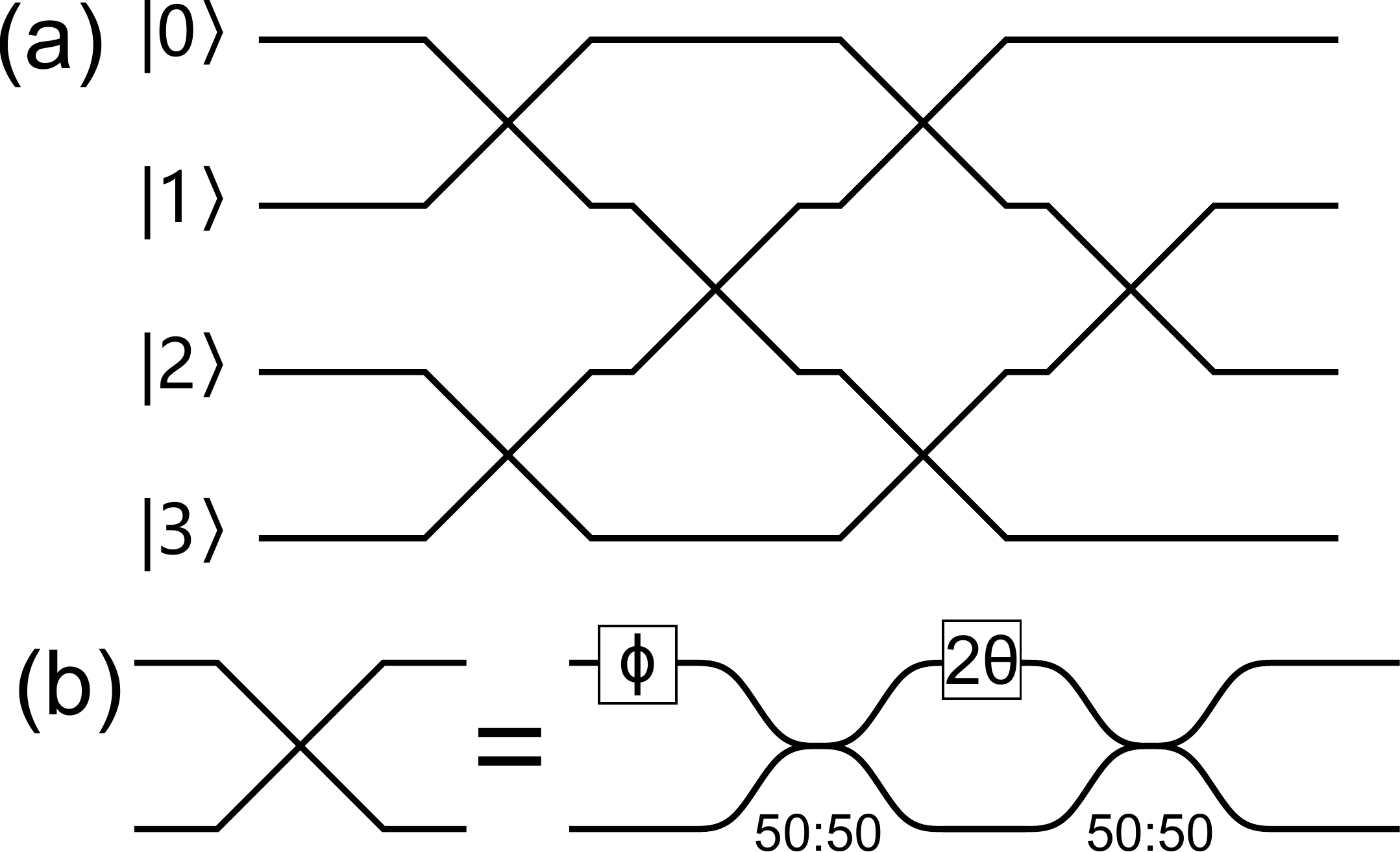}
\caption{(a) A $4\times 4$ unitary transformation can be realized by 6 variable beam splitters. Each black line represents one input mode. The crossing between two lines represents one variable beam splitter. (b) A variable beam splitter can be constructed from an MZI with two tunable phase shifters.}
\label{fig1}
\end{figure}

\section{Elimination-based decomposition}
In this section, we briefly review the rectangular design proposed by Clements \textit{et al.} in \cite{Clements:16}. The algorithm in \cite{Clements:16} eliminates the off-diagonal entries of an HD unitary transformation matrix by multiplying a transformation matrix between two adjacent modes. The residual diagonal matrix turns out to be an HD phase gate. The number of required MZIs scales as $N(N-1)/2$, where $N$ is the dimension of the quantum space in study. Figure~\ref{fig1}a shows an example of realizing a 4 $\times$ 4 unitary transformation $U_{4}$ with 6 MZIs for photonic paths. Here we only focus on the MZIs and omit the residual phase gate, since it can be realized by inserting phase shifters in each mode in the end. The input modes are encoded with the logical states $|0\rangle, |1\rangle, |2\rangle$ and $|3\rangle$, respectively, which are represented by black lines. Each crossing between two modes represents a variable beam splitter operating on the two modes, which can be realized by an MZI with two phase shifters, as shown in Fig.~\ref{fig1}b.

The decomposition of $U_{4}$ thus has the following form
\begin{equation}
U_4=D_4^{(7)}T^{(6)}_{1,2}T^{(5)}_{2,3}T^{(4)}_{0,1}T^{(3)}_{1,2}T^{(2)}_{2,3}T^{(1)}_{0,1},
\label{u4}
\end{equation}
where $T_{m,n}$ is the splitting operation between two adjacent modes $m$ and $n$ ($n=m+1$), $D_4$ is a diagonal matrix which corresponds to a phase gate, and the superscripts indicate the operating order. The matrix $T_{m,n}$ is the identity matrix with the entries at the intersection of the $m$th and $n$th rows and columns replaced by the following matrix
\begin{eqnarray}
R_{m,n}(\theta,\phi)&=&\left[\begin{array}{cc}
e^{i\phi}\cos\theta & -\sin\theta \\
e^{i\phi}\sin\theta & \cos\theta\\
\end{array}\right].
\end{eqnarray}
Note that the $\theta$ and $\phi$ of each $T_{m,n}$ in Eq.~(\ref{u4}) are not necessarily equal. For simplicity, we write $T_{m,n}$, without explicitly indicating the parameters $\theta$ and $\phi$, to represent a general splitting operation between the $m$th mode and the $n$th mode.

\section{Results}
In this section, we first present a scheme which utilizes the path and OAM DoFs of a single photon to equivalently realize the network shown in Fig.~\ref{fig1}a. We show that such a hybrid encoding reduces the number of required MZIs while maintaining the symmetric structure. Though we use the OAM DoF to encode, our scheme does not require complex operations. Only MZI structures and dove prisms are needed. Then we generalize our scheme to arbitrary HD quantum systems. 
%Finally, we show how to realize universal unitary transformations on OAM DoF based on our scheme, if an interface between path and OAM is applied.

\subsection{Path-OAM-encoding}
%Now we show how a hybrid encoding which uses the path and the OAM DoFs of a photon, can simplify the implementation.
We use two path states and two OAM states to constitute a 4-dimensional space. The quantum state in the hybrid space is $|p,\ell\rangle$, where $p$ is the path, and $\ell$ is the OAM number, with $p\in\{0,1\}$ and $\ell\in\{-1,1\}$. The logical states are encoded in the following order
\begin{eqnarray}
\nonumber  |0\rangle&\equiv&|0,1\rangle,\quad |1\rangle\equiv|1,1\rangle,\\
|2\rangle&\equiv& |0,-1\rangle,\quad|3\rangle\equiv |1,-1\rangle .
\label{encode}
\end{eqnarray}
Such encoding of the logical states will benefit the realization of the unitary transformation, which can be seen later. 

\begin{figure}[tbp!]
\centering\includegraphics[width=0.45\textwidth]{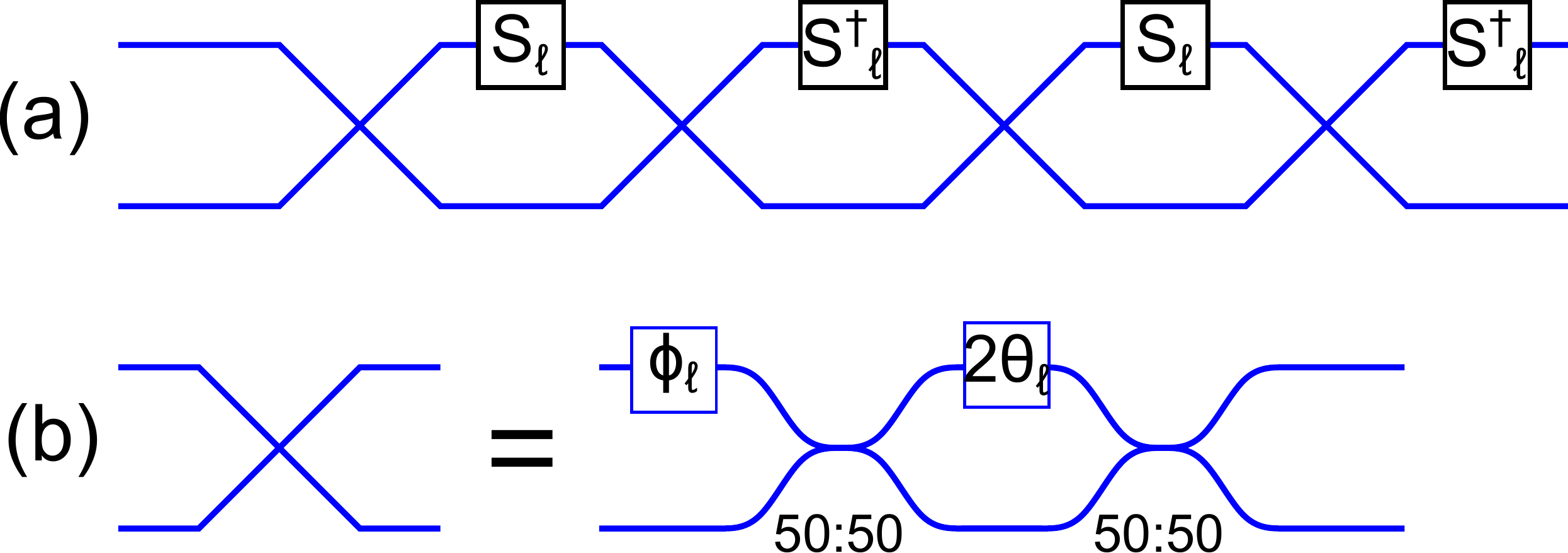}
\caption{(a) A general unitary transformation in the path-OAM hybrid space can be realized by four OAM-dependent beam splitters. Each blue line represents one path which contains two different OAM states. The crossing between two paths represents an OAM-dependent beam splitter. Two SWAP gates $S_{\ell}$ which affect the OAM DoF, are inserted in the upper path before the second and the fourth OAM-dependent beam splitters; while the two corresponding inverse operations $S^{\dagger}_{\ell}$ are placed in the upper path after the second and the fourth OAM-dependent beam splitters. (b) The OAM-dependent beam splitter can be constructed from an MZI with two phase shifters which are dependent on the OAM. For the detailed construction of an OAM-dependent beam splitter, see Supplementary Note 1.}
\label{fig2}
\end{figure}

Figure~\ref{fig2}a shows our proposed scheme to implement a 4 $\times$ 4 unitary transformation $U_4$ in the path-OAM hybrid space. The key module in our design is the OAM-dependent beam splitter, as shown in Fig.~\ref{fig2}b. It is well known that a variable beam splitter operating on the path DoF mixes two paths, which is independent of the polarization, OAM or any other DoFs of the photon. Suppose the phases $\phi$ and $\theta$ in Fig.~\ref{fig2}b are OAM-dependent, which can be realized by dove prisms, the resulting splitting operation on the path DoF will naturally be OAM-dependent (See Supplementary Note 1 for the details on the construction of an OAM-dependent beam splitter). We denote such an OAM-dependent beam splitter by $T_{\ell}^{m,n}=T_{\ell}^{m,n}(\theta_{\ell}, \phi_{\ell})$, which represents an $\ell$-dependent transformation on the $m$th path and the $n$th path. 

It is worthy to note that in such a hybrid encoding, there are two modes (two OAM states $|\pm\ell\rangle$ with $\ell=1$) in one path. Hereafter, we use a blue line to represent one path containing the two OAM states in order to distinguish it from a black line representing one mode (one path) in Fig.~\ref{fig1}a. The superscripts $m$ and $n$ in $T_{\ell}^{m,n}$ mean that the splitting operation affects paths $m$ and $n$, while subscripts $m$ and $n$ in $T_{m,n}$ of Eq.~(\ref{u4}) mean that the splitting operation affects modes $m$ and $n$. To clarify the scheme shown in Fig.~\ref{fig2}a, in the following, we explain the realizations of (i) $T_{2,3}T_{0,1}$ and (ii) $T_{1,2}$.

(i) From the layout of the MZIs in Fig.~\ref{fig1}a, one can see that $T^{(1)}_{0,1}$ commutes with $T^{(2)}_{2,3}$, so do $T^{(4)}_{0,1}$ and $T^{(5)}_{2,3}$. Here, the superscript in the parentheses is only used to refer to the specific operation in Eq.~(\ref{u4}). Note that on the basis of the encoding of the logical states in Eq.~(\ref{encode}), $T_{0,1}$ corresponds to the splitting operation between the states $|0,1\rangle$ and $|1,1\rangle$, while $T_{2,3}$ corresponds to the splitting operation between the states $|0,-1\rangle$ and $|1,-1\rangle$. Therefore, $T_{0,1}$ and $T_{2,3}$ can be regarded as OAM-dependent splitting operations on paths 0 and 1 in our scheme, with $T_{0,1}=T^{0,1}_1$ and $T_{2,3}=T^{0,1}_{-1}$. They can be simultaneously realized with a single OAM-dependent beam splitter at the cost of some additional dove prisms. 
%In Fig.~\ref{fig2}a, the first OAM-dependent beam splitter M1 functions as $T^{(2)}_{2,3}T^{(1)}_{0,1}$ in Eq.~(\ref{u4}), while M3 functions as $T^{(5)}_{2,3}T^{(4)}_{0,1}$ in Eq.~(\ref{u4}).

(ii) On the other hand, $T_{1,2}$ is the splitting operation between the states $|1,1\rangle$ and $|0,-1\rangle$, which mixes different OAMs in different paths. To avoid complex operations on the OAM states, a SWAP gate $S_{\ell}$,  which affects the OAM states, is inserted in path 0. The SWAP gate transforms the path-OAM coupled states in the following way
\begin{eqnarray}
\nonumber &&|0,1\rangle\rightarrow |0,-1\rangle,\quad |1,1\rangle\rightarrow |1,1\rangle,\\
&&|0,-1\rangle\rightarrow |0,1\rangle,\quad |1,-1\rangle\rightarrow |1,-1\rangle.
\label{swap}
\end{eqnarray}
It turns out that $T_{1,2}$, which originally represents the operation on the states $|1,1\rangle$ and $|0,-1\rangle$, is now actually the operation on the states $|1,1\rangle$ and $|0,1\rangle$. This operation can be realized by the operator $T^{0,1}_{1}$, which corresponds to an OAM-dependent beam splitter operating on paths 0 and 1 for $\ell=1$.

After the SWAP gate $S_{\ell}$, the other two states become $|1,-1\rangle$ and $|0,-1\rangle$, and they remain unchanged while $T_{1,2}$ is taking effect. Nevertheless, one can imagine an identity operation on them. The identity operation can be realized by $T^{0,1}_{-1}=\mathbb{I}$, which corresponds to an OAM-dependent beam splitter operating on paths 0 and 1, and it changes nothing for $\ell=-1$. It should be noted that the imagined identity operator here simplifies the  realization of $T_{1,2}$. The reason for this is as follows. If one only considers the two states affected by $T_{1,2}$, an OAM sorter will be required to separate the two OAMs in the two paths. Likewise, an OAM combiner is also needed to combine the OAMs after $T_{1,2}$. In this case, more complex operations on the OAM states will be required.

\begin{table}[htbp!]%The best place to locate the table environment is directly after its first reference in text
\caption{\label{table1}Correspondence between the path-encoding scheme and the path-OAM encoding scheme for realizing an arbitrary 4-dimensional transformation.}
\begin{ruledtabular}
\begin{tabular}{ccccc}
\textrm{Path-encoding}&$T^{(2)}_{2,3}T^{(1)}_{0,1}$&$T^{(3)}_{1,2}$&$T^{(5)}_{2,3}T^{(4)}_{0,1}$&$T^{(6)}_{1,2}$\\
\colrule
\textrm{Path-OAM-encoding}&$T^{0,1}_{\ell}$&$S^{\dagger}_{\ell}T^{0,1}_{\ell}S_{\ell}$&$T^{0,1}_{\ell}$&$S^{\dagger}_{\ell}T^{0,1}_{\ell}S_{\ell}$\\
\end{tabular}
\end{ruledtabular}
\end{table}

Finally, an inverse operation $S_{\ell}^{\dagger}$ of $S_{\ell}$ in path 0 is required to restore the OAM state. The SWAP gate and its inverse operation $S_{\ell}^{\dagger}$ can be readily realized by dove prisms, owing to the choice of the OAM states. Therefore, the $T_{1,2}$ in Eq.~(\ref{u4}) for the path-encoding scheme can be replaced by an OAM-dependent beam splitter $T^{0,1}_{\ell}$ sandwiched between $S_{\ell}$ and $S^{\dagger}_{\ell}$ for the path-OAM encoding scheme. 

Table \ref{table1} summarizes the correspondence between the realizations of $U_4$ based on the path-encoding scheme (Fig.~\ref{fig1}a) and the path-OAM encoding scheme (Fig.~\ref{fig2}a). To sum up, for a path-encoding scheme, one needs 6 MZIs, while for a path-OAM encoding scheme, only 4 MZI structures are required. The advantage of this hybrid encoding becomes apparent with increasing dimensionality, as is seen in the following subsections.

\subsection{Path-OAM-encoding in HD spaces}
\begin{figure}[htbp!]
\centering\includegraphics[width=0.5\textwidth]{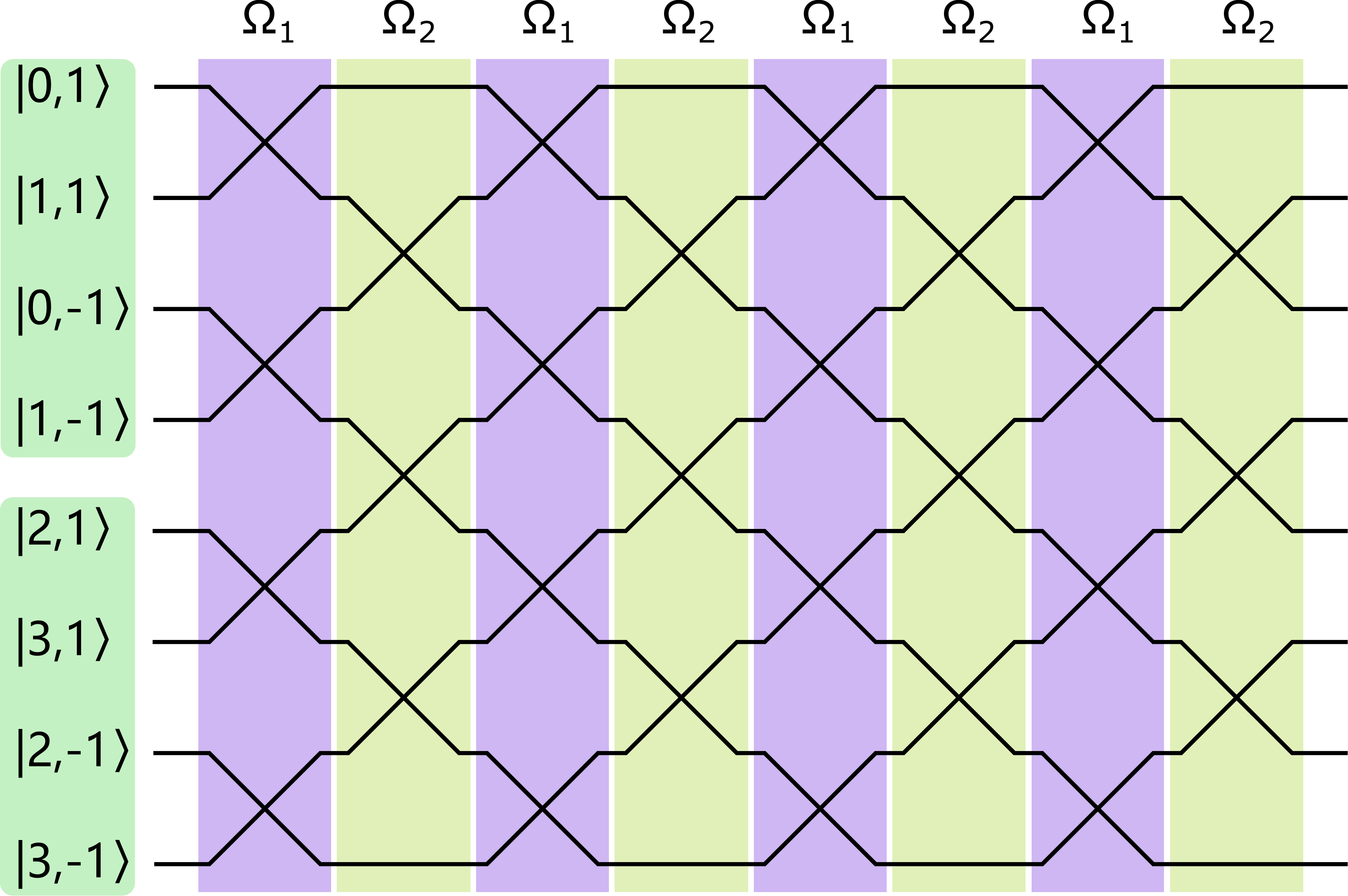}
\caption{Decomposition of an $8\times 8$ unitary transformation. The two green regions on the left denotes two 4-dimensional subspaces in the path-OAM hybrid space, each of which is formed by two path states and two OAM states. The eight input modes are encoded by the corresponding eight path-OAM coupled states. The operations in the two layers $\Omega_1$ and $\Omega_2$ act alternately on the input states.}
\label{fig3}
\end{figure}

Our scheme can be generalized to arbitrary HD quantum systems. For simplicity, we consider the realization of a $4n\times 4n$ unitary transformation, where $n$ is an integer. We still use the two OAM states ($|1\rangle$ and $|-1\rangle$) and $2n$ different path states ($|0\rangle,|1\rangle,\cdots,|2n-1\rangle$) to constitute the $4n$-dimensional space. The $4n$-dimensional space can be divided into $n$ 4-dimensional subspaces, each of which is formed by the two OAM states and two adjacent path states. To be specific, the $k$th subspace is formed by two path states $|2k-2\rangle$ and $|2k-1\rangle$, as well as the two OAM states $|\pm 1\rangle$, with $k=1,2,\cdots,n$. The basis states in the $k$th subspace are therefore $\{|2k-2,1\rangle, |2k-1,1\rangle, |2k-2,-1\rangle, |2k-1,-1\rangle\}$, and they are encoded with the logical states $|4k-4\rangle, |4k-3\rangle, |4k-2\rangle$ and $|4k-1\rangle$, respectively. The logical states are thus encoded in an order similar to Eq.~(\ref{encode})
\begin{eqnarray}
\nonumber &&|0\rangle\equiv |0,1\rangle,\quad |1\rangle\equiv|1,1\rangle, \quad  |2\rangle\equiv|0,-1\rangle,\quad |3\rangle\equiv |1,-1\rangle,\\
\nonumber &&|4\rangle\equiv |2,1\rangle,\quad |5\rangle\equiv|3,1\rangle,\quad  |6\rangle\equiv |2,-1\rangle,\quad |7\rangle\equiv |3,-1\rangle,\\
\nonumber &&\vdots\\
\nonumber &&|4n-4\rangle\equiv |2n-2,1\rangle,\quad |4n-3\rangle\equiv|2n-1,1\rangle, \\
&&|4n-2\rangle\equiv |2n-2,-1\rangle,\quad |4n-1\rangle\equiv |2n-1,-1\rangle.
\label{encodeHD}
\end{eqnarray}

The rectangular design in \cite{Clements:16} can be regarded as consisting of two different layers. One layer, denoted by $\Omega_1$, realizes the splitting operation between the $j$th mode and the $(j+1)$th mode, where $j$ is an even number; while the other layer, denoted by $\Omega_2$, realizes the splitting operation between the $(j-1)$th mode and the $j$th mode. The operations in these two layers act on the input states alternately. Figure \ref{fig3} shows the network of the rectangular design for implementing an $8\times 8$ unitary transformation based on the path encoding. On the left of Fig.~\ref{fig3}, the four path-OAM coupled states in the upper green region form a subspace, while the four path-OAM coupled states in the lower green region form another subspace.

From the encoding of the logical states in Eq.~(\ref{encodeHD}), one can straightforwardly conclude that the operations in the layer $\Omega_1$ can be implemented by OAM-dependent beam splitters operating on each 4-dimensional subspace. To be specific, for the $k$th subspace, one can realize the operations in the layer $\Omega_1$ by the OAM-dependent beam splitter $T^{2(k-1),2k-1}_{\ell}$, which operates on $2(k-1)$th and $(2k-1)$th paths. 

\begin{figure*}[htbp]
\centering\includegraphics[width=1\textwidth]{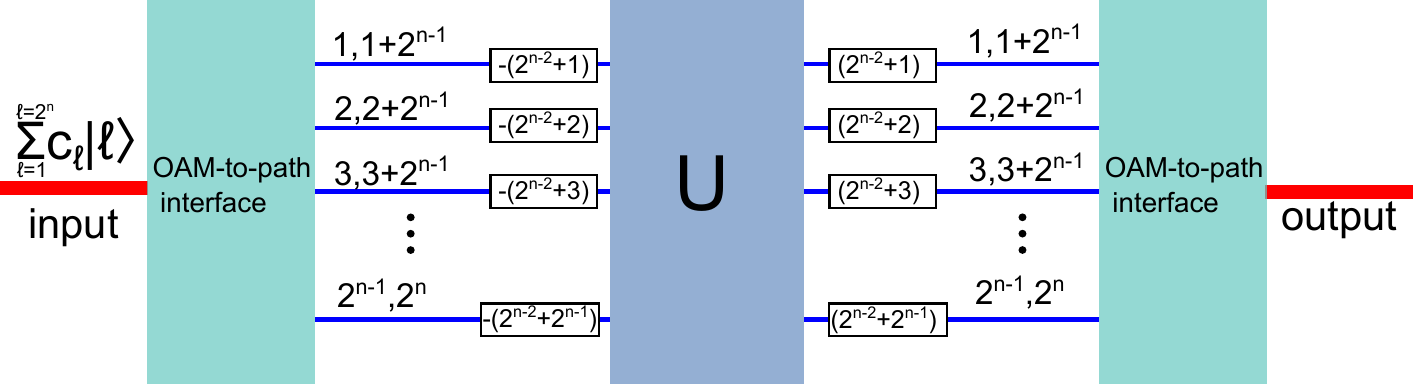}
\caption{Implementation of transformations on the OAM states. The input state is a superposition of OAM states in $2^n$-dimensional space. The OAM-to-path interface redirects different OAM states onto different paths. Each path contains two OAM states, of which the OAM numbers differ by $2^{n-1}$. Then in each path, a spiral phase plate is inserted to transform the two OAM states to $|\pm2^{n-2}\rangle$. After that, a unit (labelled by ``U'') is constructed to realize the desired unitary transformation. At the output of the unit, the spiral phase plate in each path transforms the OAM states back. Finally, a second OAM-to-path interface combines the $2^{n-1}$ paths to obtain the output state.}
\label{fig4}
\end{figure*}

On the other hand, the operations in the layer $\Omega_2$ mixes different OAMs in different paths, as shown in Fig.~\ref{fig3}. It can be realized in the same manner as Fig.~\ref{fig2}, i.e., by an OAM-dependent beam splitter and a SWAP gate. However, each path is mixed with the previous and the later paths in the layer $\Omega_2$, except the first and the last paths. To be specific, as can be seen from Fig.~\ref{fig3}, in the layer $\Omega_2$, the states $|1,1\rangle$ and $|0,-1\rangle$ are mixed ($T_{1,2}$), while the states $|1,-1\rangle$ and $|2,1\rangle$ are mixed ($T_{3,4}$). That is, path 1 is mixed with paths 0 and 2. To avoid complex OAM operations, we could first realize $T_{1,2}$ and then $T_{3,4}$. This gives rise to a longer optical depth, which is defined as the maximum number of MZIs that the photon traverses \cite{Clements:16}. In Supplementary Note 2, we present an example of implementing a $12\times 12$ unitary transformation.

For other dimensions, namely, $4n-1$, $4n-2$ and $4n-3$ dimensions, a clever way to implement an arbitrary unitary transformation is to expand the space to a $4n$-dimensional one. That is, $\mathcal{H}_{4n}:=\mathcal{H}_o\oplus\mathcal{H}'$, where $\mathcal{H}_o$ is the original space, and the dimension of $\mathcal{H}'$ is 3, 2, or 1. The corresponding quantum state and transformation matrix are mapped by $\rho:=\rho_o\oplus \mathbf{0}$ and $U:=U_o\oplus \mathbf{1}$, respectively. Here, $\rho_o$ ($U_o$) is the density matrix (transformation matrix) in the original space, and $\mathbf{0}$ ($\mathbf{1}$) is the zero (identity) operator in $\mathcal{H}'$.

\subsection{Scaling properties}
The benefit of combining path and OAM DoFs of a single photon is obvious. The previous theoretical work \cite{PhysRevA.92.043813} showed that by using path and internal DoFs of a single photon, the number of required MZIs could be reduced by a factor $n^2_p/2$, where $n_p$ is the dimension of the internal DoF, at the cost of more optical elements acting on the internal DoF. Nevertheless, the conclusion could only be suitable when the internal DoF is photonic polarization, since operations on the photonic polarization can be realized by combinations of wave plates without need of MZIs. If other internal DoFs are applied (e.g., OAM in our work), the conclusion would no longer be tenable, since the manipulation of the OAM states generally requires more MZIs. 

Here we note that using two OAM states ($|\pm\ell\rangle$) coupled with the path DoF, the number of required MZIs can still be reduced. As shown in Supplementary Note 2, each layer $\Omega_1$ requires half number of MZIs due to the usage of OAM-dependent beam splitters. For dimensionality of $4n$, the number of required MZIs scales as  $2n(4n-1)$ for the scheme in \cite{Clements:16}. In our proposal, the operations in the two layers $\Omega_1$ and $\Omega_2$ are performed alternatively for $2n$ times. Thus, the total number of required MZIs is reduced by $2n^2$, i.e., about one quarter of the total. Therefore, the number of required OAM-dependent beam splitter scales as $6n^2-2n$, and the number of the $S_{\ell}$ and $S^{\dagger}_{\ell}$ gates is $4n^2$. However, the optical depth of our scheme is $6n$ which is larger than that of \cite{Clements:16} since the operations in the layer $\Omega_2$ are realized in two steps (See Supplementary Note 2 for details).

\section{Universal unitary transformations on the OAM states}
A universal implementation of unitary transformations on the OAM states is important in OAM-based applications. In the previous sections, though we use two OAM states to couple the path DoF to constitute an HD space, we illustrate here that our scheme can be adopted to realize unitary transformations on the OAM states alone.

For simplicity, consider a $2^n$-dimensional HD space, which is formed by OAM states $\{|\ell\rangle\}$ with $\ell=1, 2, \cdots, 2^n$. The network to implement an arbitrary unitary transformation on the OAM states is shown in Fig.~\ref{fig4}. The input state is a superposition of $2^n$ OAM states. Then an OAM-to-path interface redirects the OAM states onto $2^{n-1}$ paths. Each path contains two OAM states. The OAM states $|1\rangle$ and $|1+2^{n-1}\rangle$ are redirected onto path 0, the OAM states $|2\rangle$ and $|2+2^{n-1}\rangle$ are redirected onto path 1, etc. The OAM-to-path interface can be constructed from cascaded OAM sorters (See Supplementary Note 3 for details) or a mode sorter \cite{fickler2014interface}. After that, each path is inserted with a spiral phase plate which decreases the OAM number by $(2^{n-2}+i)$, where $i$ is the path number with $i=0,1,\cdots,(2^{n-1}-1)$. The spiral phase plates transform each pair of OAM states in the $2^{n-1}$ paths to the states $|\pm 2^{n-2}\rangle$, i.e., the same magnitude but opposite signs. At this step, the $2^n$-dimensional space formed by $2^n$ OAM states is mapped to a $2^n$-dimensional space formed by $2^{n-1}$ path states and the two OAM states $|\pm 2^{n-2}\rangle$. 

Then a unit, which can be constructed according to the previous subsection, realizes the desired unitary transformation on the path-OAM coupled state. Finally, each path is inserted with another spiral phase plate which increases the OAM number by $(2^{n-2}+i)$ to transform the OAM states back. A second OAM-to-path interface combines the $2^{n-1}$ paths to obtain the output state. The optical elements used in our scheme for the universal implementation of arbitrary transformations on an OAM state in $2^n$-dimensional space are summarized in table \ref{table2}. 

\begin{table}[htbp!]%The best place to locate the table environment is directly after its first reference in text
\caption{\label{table2}Summary of the optical elements used for implementing arbitrary unitary transformations on OAM states in $2^n$-dimensional space. SPP: spiral phase plate, OBS: OAM-dependent beam splitter.}
\begin{ruledtabular}
\begin{tabular}{cccc}
\textrm{SPP}&\textrm{OAM sorter}&\textrm{$S_{\ell}$ and $S^{\dagger}_{\ell}$}&\textrm{OBS}\\
\colrule
$2^n$ & $2^n-2$ & $2^{2n-2}$ & $3\cdot 2^{2n-3}-2^{n-1}$\\
\end{tabular}
\end{ruledtabular}
\end{table}

\section{Conclusion}
We have presented an efficient scheme which realizes arbitrary unitary transformations on quantum states encoded by the path and OAM DoFs of a single photon. By using OAM-dependent beam splitters, our scheme incorporates two splitting operations into a single operation of an OAM-dependent beam splitter. Thus, our scheme reduces the number of required MZIs while maintaining the symmetric architecture proposed in \cite{Clements:16}, at the cost of additional dove prisms and longer optical path. In principle, our scheme can be generalized to path-OAM encoded quantum states with the subspace of OAM DoF greater than 2. However, if more OAM states are considered in the hybrid space, implementing the transformations will become complicated because it is difficult to perform operations on arbitrary OAM states.

Furthermore, we have demonstrated that a universal implementation of transformations on the OAM states can be constructed based on our scheme. With additional OAM-to-path interfaces and spiral phase plates, our proposal provides an explicit realization of transformations on the OAM states. While it could be experimentally challenging to realize our scheme in high dimensions, the optical elements used are all commercially available. One could expect that our scheme is realizable in bulk optics with low dimensions, or in integrated optics with the advance of the technique. Our work fills in the gap regarding arbitrary transformations on photonic OAM states and it could have potential applications in OAM-based HD quantum communication and HD quantum computation.

Finally, we note that two OAM states can be equivalent to a polarization DoF. However, to manipulate different DoFs requires different optical elements. For example, manipulating polarization generally requires wave plates; while manipulating OAM requires different optical elements, such as spatial light modulators, dove prisms, or spiral phase plates, etc. The key point of our scheme is that we maintain the symmetric layout proposed in Ref. \cite{Clements:16}, which tolerates the systematic errors. Therefore, our scheme not only reduces the required interferometers but also is robust to the noise. Another advantage is that, as demonstrated in section 4, our scheme can be generalized to realizing unitary operations on the OAM states, which is a vital issue in OAM-based HD quantum computation.

\begin{acknowledgments}
This work was partly supported by the National Natural Science Foundation of China (NSFC) (12204312, U21A20436, 11774076), Jiangxi Provincial Natural Science Foundation (20224BAB211014), and the Key-Area Research and Development Program of GuangDong province (2018B030326001).
\end{acknowledgments}

~\\
\noindent\textbf{Competing interests:}\\
The authors declare no competing interests\\
~\\
\noindent\textbf{Data and materials availability:}\\
The data that support the findings of this study are available from the corresponding
authors upon reasonable request.\\

%\clearpage
%\newpage
\appendix

\onecolumngrid
%\begin{widetext}

\section{Supplementary Note-1: Construction of an OAM-dependent beam splitter}
\begin{figure}[htbp!]
\centering\includegraphics[width=0.55\textwidth]{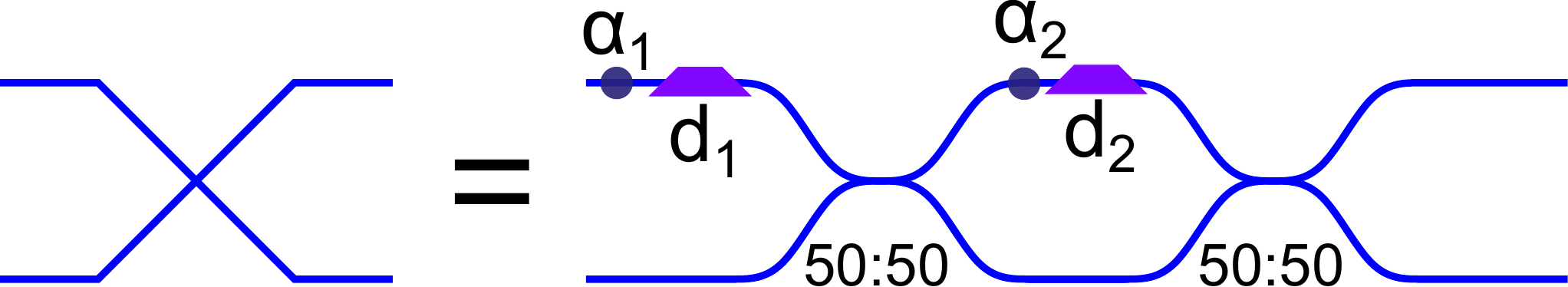}
\caption{Construction of an OAM-dependent beam splitter. Each blue line represents one path containing two OAM states. $\alpha_1$ and $\alpha_2$ are the phases introduced by two phase shifters. The two dove prisms \textrm{d}$_1$ and \textrm{d}$_2$ introduce OAM-dependent phases.}
\label{figs1}
\end{figure}

One possible construction of an OAM-dependent beam splitter is shown in Fig.~\ref{figs1}. The phase shifter (represented by a dot) before the MZI introduces a phase $\alpha_1$ to the upper path, followed by a dove prism (\textrm{d}$_1$) oriented at $\beta_1$. Here, a blue line represents one path containing two OAM states. The dove prism introduces an OAM-dependent phase $2\beta_1\ell$ to the photon carrying OAM, with an additional reflection. Note that here we neglect the reflection caused by the dove prism, as it can be eliminated by inserting one more dove prism  or another mirror. Thus, the phase appended to the upper path is $\phi=\alpha_1+2\beta_1\ell$. Similarly, inside the MZI, another phase shifter and dove prism introduce phase $2\theta=\alpha_2+2\beta_2\ell$, where $\beta_2$ is the orientation of the second dove prism.

Suppose the desired splitting operations on the two OAM states $\ell=1$ and $\ell=-1$ are $U_1(\phi_1, \theta_1)$ and $U_{-1}(\phi_{-1}, \theta_{-1})$, respectively. Here, $\phi_i$ and $\theta_i$ ($i=1, -1$) are the angles which need to be set by the phase shifters and dove prisms, to realize the desired operation $U_i$. By straightforward calculation, one has $\alpha_1=\frac{\phi_1+\phi_{-1}}{2}$, $\beta_1=\frac{\phi_1-\phi_{-1}}{4}$, $\alpha_2=\frac{\theta_1+\theta_{-1}}{2}$, $\beta_2=\frac{\theta_1-\theta_{-1}}{4}$. Therefore, such a structure can realize two arbitrary splitter operations respectively on the two OAM states $\ell=1$ and $\ell=-1$.

Note that such a construction can only realize two different splitting operations on two different OAM states, since the phases introduced by the dove prisms depend linearly on the OAM number. If a third OAM state, with the OAM number different from the other two, is input, the splitting operation on the third OAM state could be different from those on the other two states, but it is dependent on those two operations.

\section{Supplementary Note-2: Implementation of $12\times 12$ unitary transformations}
\begin{figure*}[htbp!]
\centering\includegraphics[width=1\textwidth]{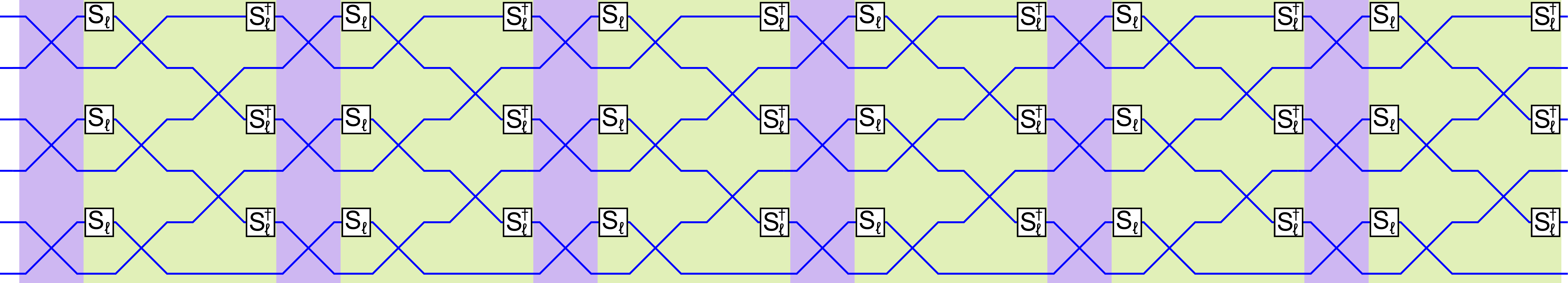}
\caption{Implementation of a $12\times 12$ unitary transformation. The operations in the two layers, $\Omega_1$ (represented by the purple region) and $\Omega_2$ (represented by the greenish yellow region), are performed alternately for six times. The realization of $\Omega_2$ requires a SWAP gate and its inverse operation which act on the OAM DoF.}
\label{figs2}
\end{figure*}

\begin{figure}[htbp!]
\centering\includegraphics[width=0.55\textwidth]{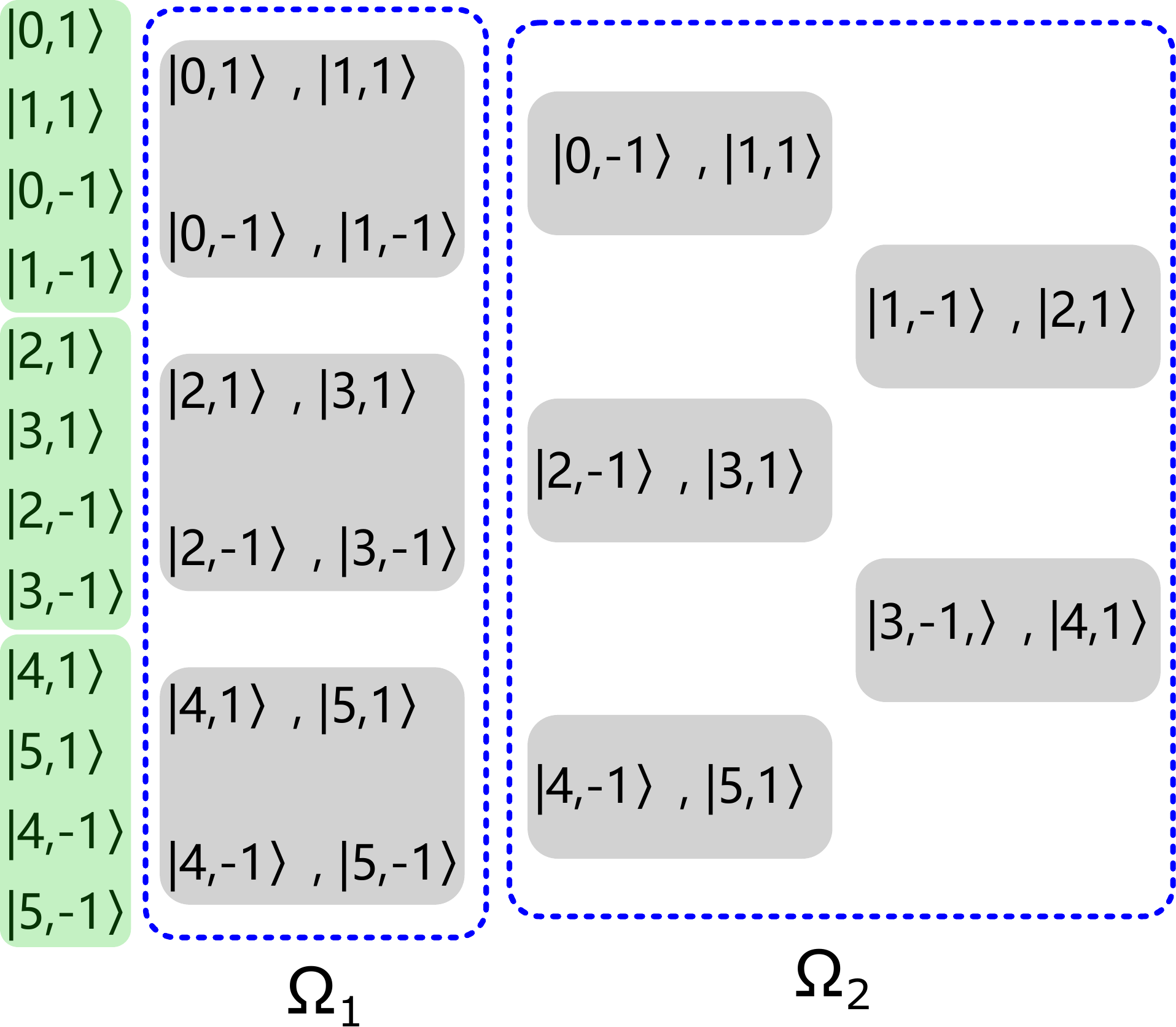}
\caption{The path-OAM coupled states in the green regions are used to encoded the logical states $|0\rangle \sim |11\rangle$. The layer $\Omega_1$ contains three pairs of splitting operations, each of which can be realized by an OAM-dependent beam splitter. }
\label{figs3}
\end{figure}

Figure \ref{figs2} shows the implementation of a $12\times 12$ unitary transformation based on our proposed path-OAM encoding scheme. The two OAM states are coupled with six path states to constitute a 12-dimensional space. Each blue line represents one path containing two OAM states. Figure \ref{figs3} shows the corresponding quantum states. In Fig.~\ref{figs3}, the three green regions indicate three subspaces. The logical states, $|0\rangle \sim |11\rangle$, are encoded by the path-OAM coupled states in the green regions from top to bottom. The gray regions in the dashed boxes indicate the operations in $\Omega_1$ and $\Omega_2$ layers. Each gray region represents an OAM-dependent beam splitter in our scheme.

As shown in Figs.~\ref{figs2} and \ref{figs3}, the two layers, $\Omega_1$ and $\Omega_2$, operate on the input state alternatively. Each $\Omega_1$ layer is realized by three OAM-dependent beam splitters, i.e., $T_{\ell}^{0,1}$, $T_{\ell}^{2,3}$ and $T_{\ell}^{4,5}$, shown by the three gray regions in Fig.~\ref{figs3}. On the other hand, the operations in each $\Omega_2$ layer act on five pairs of OAM states, shown by the five gray regions in Fig.~\ref{figs3}. The realization of each $\Omega_2$ layer is divided into two steps. In the first step, we realize the three operations shown by the left three gray regions inside the right dashed-line block in Fig.~\ref{figs3}. To realize the splitting operation between the states $|0,-1\rangle$ and $|1,1\rangle$, an $S_{\ell}$ gate, which acts on the OAM DoF, is inserted in path 0. After that, an OAM-dependent beam splitter suffices to realize the desired operation. Then, an $S^{\dagger}_{\ell}$ gate is inserted to transform the OAM states back. Likewise, here we imagine an identity operator acting on the other two states in the first subspace. This procedure applies to the realizations of the splitting operation between the states $|2,-1\rangle$ and $|3,1\rangle$, and the splitting operation between the states $|4,-1\rangle$ and $|5,1\rangle$. 

In the second step, we realize the other two operations shown by the right two gray regions inside the right dashed-line block in Fig.~\ref{figs3}. Note that the inverse operations of the SWAP gates in paths 2 and 4 are inserted after this step. This is straightforward since an $S_{\ell}$ gate after the first step, together with an $S^{\dagger}_{\ell}$ gate before the second step, is equivalent to the operation described by an identity operation.

\section{Supplementary Note-3: Construction of an OAM-to-path interface}
\begin{figure}[htbp!]
\centering\includegraphics[width=0.75\textwidth]{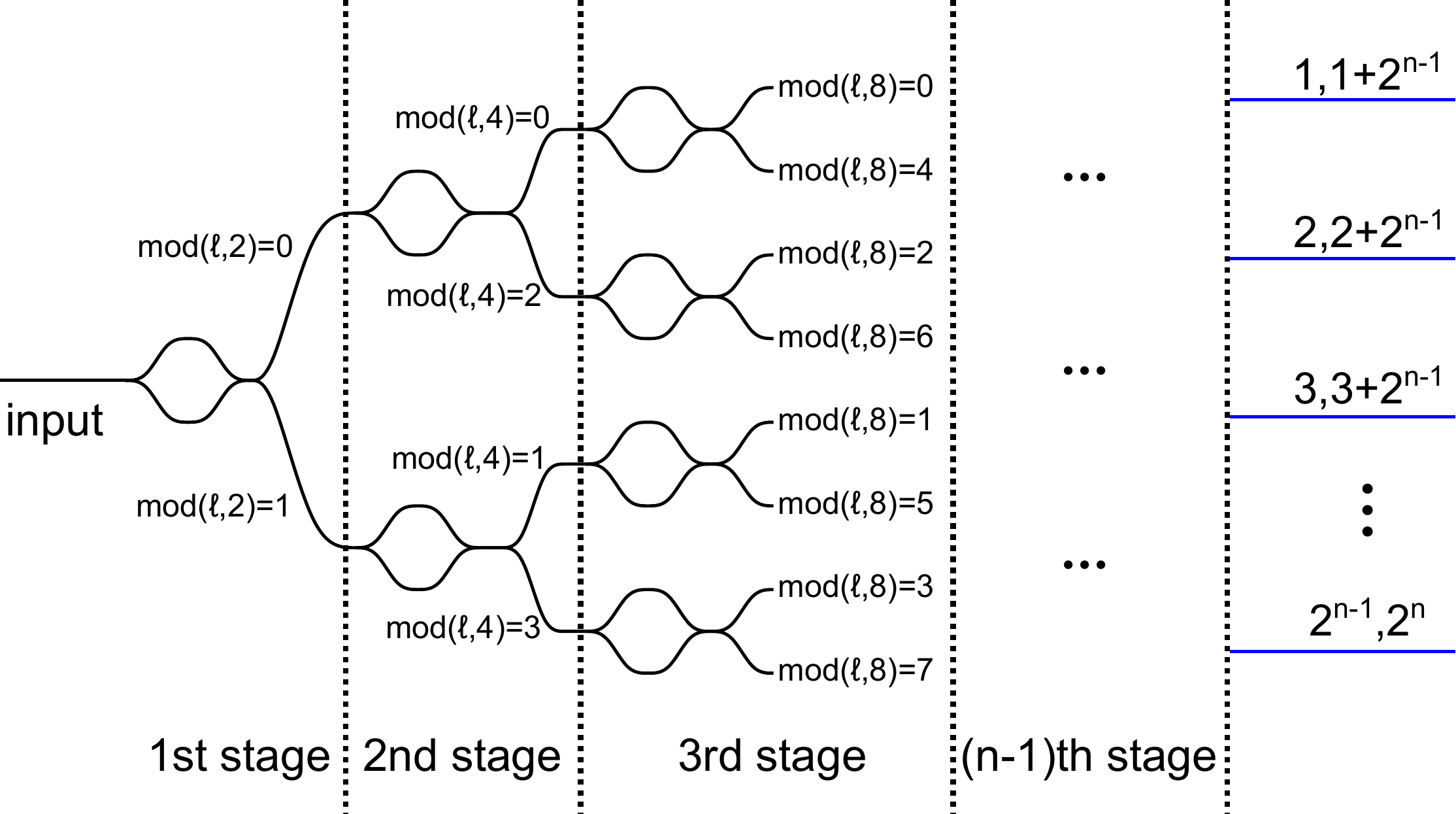}
\caption{Cascaded OAM sorters redirect different OAM states onto different paths.}
\label{figs4}
\end{figure}

One possible construction of an OAM-to-path interface is shown in Fig.~\ref{figs4}. Each MZI is equipped with two dove prisms (not shown) in two arms, known as an OAM sorter. Such an OAM sorter, when acting on the input photon carrying an OAM, can sort the OAM states according to the parity of the OAM number at the first stage, i.e., the photon with even OAM numbers goes one path, while the photon with odd OAM numbers goes the other path. A second stage, which consists of two OAM sorters, can further sort the OAM states, i.e., the photon with OAM number $\textrm{mod}(\ell,4)=0, 1, 2, \textrm{and}$ 3 is redirected onto different paths. After the $(n-1)$th stage, the OAM states are redirected onto $2^{n-1}$ paths. Each path (denoted by the blue line) contains two different OAM states, of which the OAM numbers differ by $2^{n-1}$.

\vspace{0.5cm}

%\end{large}

\twocolumngrid


\end{document}